# First Standard Quantification of Ultrasound Attenuation in Healthy Periodontal Soft Tissues *In Vivo*

Daria Poul[1,*], Amanda Rodriguez Betancourt[2], Ankita Samal[3], Carole Quesada[1], Ted Lynch[4], Cristel Baiu[5], J. Brian Fowlkes[1,6], Hsun-Liang Chan[6], Oliver D. Kripfgans[1,7,**]

[1]Department of Radiology, University of Michigan, Ann Arbor, MI, USA
[2]Department of Periodontics, College of Dentistry, University of Illinois Chicago, Chicago, IL, USA
[3]Department of Periodontics, University of Iowa College of Dentistry & Dental Clinics, Iowa City, IA, USA
[4]Sun Nuclear Corporation, Norfolk, VA, USA
[5]Department of Medical Physics, University of Wisconsin-Madison, Madison, WI, USA
[6]Department of Biomedical Engineering, University of Michigan, Ann Arbor, MI, USA
[7]Division of Periodontology, College of Dentistry, The Ohio State University, Columbus, OH, USA

*Corresponding Author 1: ssheykho@umich.edu;
**Corresponding Author 2: greentom@umich.edu;
Address: University of Michigan, Department of Radiology
6410B Med Sci I, 1301 Catherine Street, Ann Arbor, MI 48109-2026
Phone: +1 734 647 0852

## Abstract

**Objective:** Periodontal diseases affect 46% of adults aged $\geq 30$ years in the USA, yet current clinical diagnostic approaches are subjective, qualitative, and late-stage indicators. This gap highlights a critical unmet need for alternative biomarkers. Ultrasonography is emerging to fill this gap as a surrogate for non-invasive and quantitative assessments of oral diseases. This study presents the first quantifications of the ultrasound attenuation coefficient slope (ACS), as a key acoustic property and a potential biomarker of oral tissues, using standard techniques.
**Methods:** In a swine cohort (N=10), we characterized high-frequency (24 MHz) ACS of healthy periodontal tissues (gingiva) *in vivo* using the spectral difference method. First, we validated the technique using custom tissue-mimicking phantoms with known ACS. Five interproximal oral sites from each of the oral quadrants were enrolled and imaged: Premolar 3-Mesial, Premolar3-Distal, Premolar4-Distal, Molar1-Distal, and Molar2-Distal. A total of 162 oral sites were analyzed after applying exclusion criteria.
**Results:** The respective medians (1$^{st}$-quartile|3$^{rd}$-quartile) ACSs for these five oral sites were 1.66 (1.25|1.99), 1.37 (1.06|1.64), 0.99 (0.8|1.25), 1.08 (0.89|1.47), and 1.28 (0.94|1.24) dB/MHz.cm. The gingival attenuation mean at Premolar3-Mesial was significantly higher than any other oral sites ($p<0.05$) while the rest of them showed non-significance difference in their means. Across non-significant oral sites, the average ACS was 1.17 ($\pm 0.49$) dB/MHz.cm.
**Conclusion:** High-frequency ultrasound ACS of periodontal soft tissues was quantified *in vivo* using standardized techniques. This work not only characterized an important acoustic property of oral tissues for the first time but also contributes to future development of QUS biomarkers for dental healthcare that rely on attenuation knowledge.

**Keywords**: Periodontal Tissues, Ultrasound Imaging, Attenuation, Quantitative Ultrasound, Tissue Characterization, Gingiva.

# 1 Introduction

## 1.1 Overview of Periodontal Soft Tissue Structure and Periodontitis Diagnosis

In the United States, approximately 46% of adults aged 30 years or older suffer from periodontal (gum) diseases [1]. These diseases cause excessive pain that negatively impact quality of daily life. If they remain untreated, periodontal diseases can advance to tooth loss, resulting in substantial discomfort and significant financial burden from associated restorative procedures such as dental implants. Moreover, studies have also found links between periodontal diseases and systemic effects such as cardiovascular diseases [2]. Therefore, timely and accurate diagnosis, particularly at earlier stages, along with effective monitoring of treatment procedure and healing are essential for individual and public health as well as the healthcare system.

Periodontal diseases represent spectrums of chronic inflammatory conditions associated with the gum (soft) tissues that support teeth. We will first present a simple introduction of the oral soft tissue structures and the trigger region for onset of these diseases. Gum tissues, which are illustrated in detail in Figure 1 are mainly comprised of alveolar mucosal tissues and gingival tissues that are covered with a thin and highly keratinized tissue layer named epithelium. Early, reversible oral soft tissue inflammation begins from gingival tissues (termed gingivitis). Gingival soft tissues, which are densely packed with fibrous connective tissues, primarily serve as supporter of the root and bone from degradation and has two major parts. The first is free (marginal) gingiva which is a section of the gum tissues at its coronal (top) edge where tissue meets the tooth. Free gingiva is not tightly bound to the bone/tooth. The attached gingiva, on the other hand, is tightly attached to the bone and stretches to the alveolar mucosal tissues (i.e. lining mucosa), a thin tissue layer lining the bone, as shown in the illustration. The looseness of free gingiva defines a space between tooth and gum tissues, termed gingival sulcus, where it is typically flossed. Gingivitis (early oral inflammation), which results from bacteria accumulation at the sulcus, begins in the free gingiva and then spreads to the attached gingiva. Later irreversible progression of oral inflammation is termed as periodontitis. Thus, gingival tissues, as the region of early inflammation onset, serves as a potential region for exploring any early disease biomarkers.

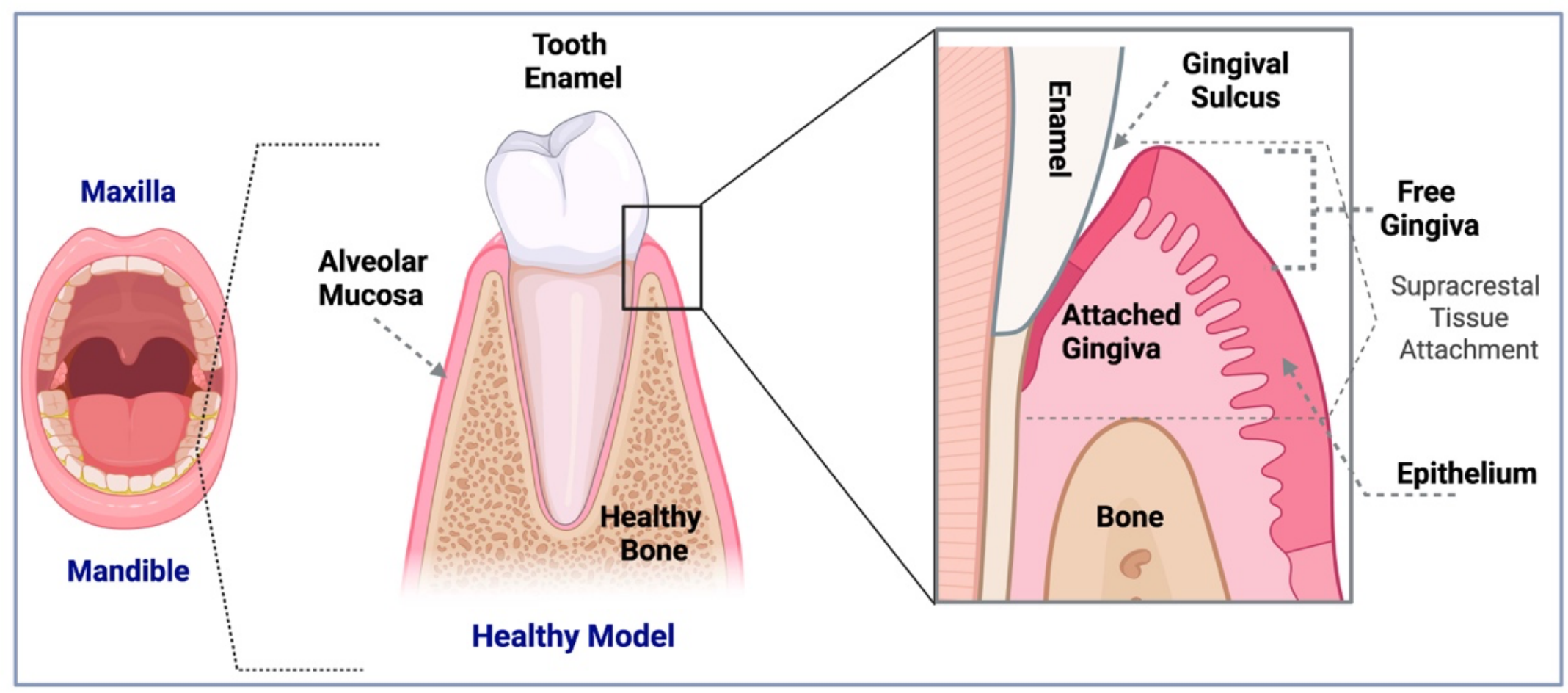

**Figure 1**. Illustration of human oral anatomy of a premolar tooth with important landmark structures annotated.

Despite the significance of such clinical needs, current tools used as standard of care in clinics have major limitations in assessing oral tissue health and diagnosing periodontal diseases objectively and quantitatively. For example, bleeding on probing (BOP) is a widely used clinical procedure for inflammation assessment [3] where a marked periodontal probe is inserted into the gingival sulcus (i.e. the oral space between tooth and gum) and the penetration depth and the presence of bleeding are recorded as indicators of inflammation extent. However, BOP is invasive, only semi-quantitative due to the probe's 1-mm discrete markings. It is also a late indicator of tissue damage, inherently subjective and highly sensitive to operator-dependent factors including the probing angle and the applied force [4]. Additionally, radiography, such as intraoral X-Ray and cone-beam CT, which is used to assess bone loss from advanced inflammation stages, expose patients to ionizing radiations while also providing limited soft tissue contrast, making them ill-suited modalities for high resolution imaging of oral soft tissues such as gingiva and lining mucosa.

## 1.2 Ultrasonography

Applications of ultrasound imaging in periodontology, as a non-ionizing, portable and cost-effective modality, have emerged rapidly over the past decade [5, 6]. Ultrasound has been increasingly explored for oral soft tissue imaging and for extracting quantitative information to characterize periodontal tissues [7-14]. While ultrasound has been a key modality in clinical diagnosis for decades in areas such as liver diseases, breast and thyroid cancers, and fetal imaging, its adoption in periodontal diagnostics has been slower. This in part originates from the smaller and more complex geometry of the oral cavity, closeness of oral tissues to highly scattering structures, such as teeth and bone, and the need for customized, smaller, high frequency ultrasound transducers for oral applications.

Ultrasound imaging could be employed as a surrogate for non-invasive and quantitative assessment of periodontal diseases. Conventional ultrasound images, namely B-mode images, with a signature grey and granular appearance, feature internal landmarks within tissues such as arteries and organ boundaries and their echogenicity (relative brightness in the B-mode image) [15, 16]. B-mode images provide a *qualitative* representation of the underlying tissue structure with certain tools for landmark measurements. As ultrasound waves penetrate tissues, they constantly interact with the tissue's underlying microstructure, returning echoes due to different acoustic properties of these microstructures. Returning echoes from tissues incorporate extensive information from interactions with tissue microstructure other than that shown on a traditional B-mode image reconstruction. Advanced quantitative assessment of these raw echoes through various statistical and mathematical techniques allows extracting additional information on underlying tissues that are otherwise hidden from a traditional B-mode representation of echoes. Quantitative ultrasound (QUS) refers to such categories of parameters obtained from ultrasound backscatter to characterize acoustic and structural properties of tissues [17]. An important QUS-based lens into tissue structure is employing ultrasound speckle statistics (spatial domain) to investigate sub-resolution tissue structures from speckle (interference) patterns, which are not typically "visible" within the imaging resolution of the ultrasound system [18, 19]. Applications of speckle statistics in tissue characterization include but not limited to human cervix [20], breast masses [21], lymph nodes of colorectal-cancer patients [22], liver diseases [23, 24], and oral soft tissues [8, 25]. Speckle statistical modeling is an active research area with novel approaches continuously evolving to expand or improve its application [26-30]. Another significant class of QUS analyses stems from spectral domain analysis, that is, transforming ultrasound echoes from the time domain to the frequency domain to analyze them. These analyses have led to extracting meaningful information representing a certain signature of tissue microstructure and scatterer properties, such as scatterer size, acoustic concentration, backscatter coefficient, and ultrasound attenuations [31-36]. The current study focuses on ultrasound attenuation quantification of oral soft tissues, which will be expanded in great details with a context from other soft biological tissues in the next section.

**Significance of Ultrasound Attenuation:**

Among QUS-derived tissue features, ultrasound attenuation as a well-established QUS parameter has significance from multiple aspects. Attenuation is a measure of ultrasound energy loss as it propagates in tissues from interactions with internal structures. As the ultrasound pulse travels within tissues, fractions of its acoustical energy are either transferred to the tissue to heat or redirected into other directions than the propagation path. These two mechanisms are known as absorption and scattering. Collectively, absorption and scattering define the tissue-specific loss mechanism.

The accurate attenuation measurement is foundational in reliable tissue ultrasound imaging, as summarized below:

- Ultrasound attenuation affects tissue imaging at deeper regions, deteriorating the image resolution/quality with depth through a decreased echogenicity. Thus, quantifying ultrasound attenuation could allow for a controlled amplification of ultrasound echoes with depth (gain compensation) to enhance image contrast at deeper regions.

- Also, attenuation measurements may lead to a unique biomarker for diagnosing the normal versus pathological conditions in biological tissues through comparing energy loss (absorption) mechanisms. Researchers have focused on attenuation characterization of healthy tissues as the necessary first step. For instance, Gray et al. [37] for the first time characterized attenuation of human pancreatic samples and discussed the impact of accurate quantification in ultrasound-based therapies. In another study, Deeba et al. [38] characterized the attenuation coefficient of human placentas in normal condition to provide a basis for future implementation of QUS techniques in disease characterization. Sebastian et al. [39] quantified the ultrasound attenuation of engineering tissues along with other acoustic properties to gain deeper insights into their structure and cellular composition. Nasief et al. [40] quantified the attenuation of breast fat as a standard for tumor attenuation comparison. Nam et al. [41] characterized attenuation coefficient of breast masses as a potential diagnostic biomarker for benign versus malignant tumors. In liver disease characterization, ultrasound attenuation has been investigated for over four decades [42]. Liver attenuation is currently implemented for fat quantification in commercial ultrasound scanners used in clinics through various propriety algorithms, as summarized in a 2022 article by the AIUM-RSNA QIBA Pulse-Echo Quantitative Ultrasound (PEQUS) initiative [43]. With such advancements in characterization of various soft tissues, the attenuation of oral soft tissues is still unknown and not investigated using any of validated standard techniques.

  It is noted that while one study in the literature has reported measurements attributed to the attenuation of oral mucosal tissues, the methodology employed deviates substantially from physical acoustics principles used to understand and accurately quantify attenuation in soft tissues in many standard techniques [46]. Specifically, the reported measurements were derived from image processing and analysis of pixel intensities of log-compressed ultrasound images rather than analyses of raw ultrasound echo signals which forms the basis for standard attenuation estimation techniques. In addition, no validations were reported to support the image-based measurements. Finally, not the actual attenuation was considered but rather the backscatter.

- In addition, accurate quantification of ultrasound attenuation is central to reduce bias in some of other QUS analyses. Particularly, for those QUS techniques that aim to extract the depth-resolved tissue scatterer information such as scatterer size or concentration, the signal loss from the attenuation mechanism needs to be accurately compensated. This is to ensure that the derived parameters reflect the tissue microstructure signature at each depth rather than the effects from the cumulative amplitude loss along depth. For example, H-scan analysis [44], which characterizes scatterer sizes, relies on accurate attenuation compensation. Christensen et al. [45] also demonstrated that the accurate ultrasound attenuation correction affects the quantification of the first order speckle statistics parameters for tissue characterization.

Thus, this highlights the crucial significance of quantifying ultrasound attenuation in periodontal tissues. This not only enables direct characterization of their attenuation mechanism, but it also serves as the foundation for reliable analyses of other QUS techniques towards advancing ultrasound in dental healthcare. In this study, we aim to quantify the *in vivo* ultrasound attenuation coefficient of oral soft tissues in a preclinical study involving a swine cohort. To the best of our knowledge, this presents the first study assessing attenuation using standard estimations techniques which were further validated by investigations on tissue-mimicking phantoms, i.e. with known acoustic properties.

# 2 Theory

Various standard techniques have been proposed in landmark research articles to quantify ultrasound attenuation in time or frequency domain representation of echoes. Time domain techniques focus on echo amplitude or energy loss of the radiofrequency signal with propagating distance [46, 47]. While time-domain techniques are easier to implement, they are more sensitive to pulse distortion, particularly high for tissues with heterogenous structures.
Frequency-based techniques, relying on the spectral power distribution of the signal, include the spectral shift method [48, 49], the spectral difference method [50, 51], the spectral log difference method [52], and hybrid method [53, 54].
In estimating ultrasound attenuation from spectral signal decay, it is important to acknowledge that the signal depth-dependent decay not only depends on tissue-specific loss mechanisms but also changes from system-related diffraction and scattering. For a more accurate estimation of attenuation as the characteristic feature of the tissue behavior, these effects should be accounted for or at best minimized. It is noted that the spectral shift method does not correct for system effects when estimating attenuation unlike the three latter spectral techniques which all employ a reference phantom with known acoustic properties to cancel out system-dependent effects.

Here, for attenuation estimation of periodontal tissues, we employed the spectral difference method as a widely used technique to eliminate system-related diffraction effect.

**Diffraction-Corrected Attenuation Estimation: Spectral Difference Method (Reference Phantom Method)**

The spectral difference method (SDM) is based on the power decay of a signal's frequency components. This technique uses a reference phantom for estimating the acoustic attenuation coefficient to eliminate system-dependent and diffraction losses towards isolating the tissue-specific attenuation. The reference phantom has known acoustic properties (speed of sound and acoustic attenuation) and is scanned using the same imaging system and image acquisition parameters as those used for tissue imaging such as depth and focus position.

To estimate attenuation within a 2D region of interest (ROI) starting at a depth of $z_0$ using SDM, each ROI scanline is split into several overlapping gates, each representing a depth band. Each time-gated RF (radiofrequency) data is converted into its power spectra, representing that depth band using the fast Fourier transform. The corresponding ROI in the reference phantom scan undergoes the same processing to ensure matched analysis between the tissue and reference phantom. The power spectrum of the time-gated RF data in a homogeneous tissue can be modeled using equation (1).

$$S(f,z) = P(f) \cdot B(f,z) \cdot D(f,z) \cdot A(f,z) \tag{1}$$

In this equation, $S(f,z)$ is the measured frequency-dependent power spectrum from tissue area at a depth of $z$. $P(f)$ reflects the joint effect of the transducer's transmit response and the transducer's sensitivity to acoustic wave energy (transfer function). $B(f,z)$ represents the tissue scattering contribution to the power spectra, while taking into account the effect of scatterer size and the number density. $D(f,z)$ models the transducer diffraction effect. Finally, $A(f,z)$ denotes the effect of tissue-specific cumulative signal loss affecting the power spectra received at the transducer, completing a round-trip between the tissue surface and the depth $z$. To isolate the local attenuation coefficient in an ROI from the cumulative attenuation that represents losses starting at tissue surface, $A(f,z)$ is defined as equation (2).

$$A(f,z) = A(f,z_0)e^{-4\alpha(f)(z-z_0)} \tag{2}$$

Here, $A(f,z_0)$ is the cumulative attenuation from the round-trip between the tissue surface and the ROI start depth $z_0$.The formulation assumes that for a given frequency, attenuation shows an exponential signal decay in a unit distance. $\alpha(f)$ is the frequency-dependent attenuation coefficient within the ROI, assuming the linear dependency of the attenuation coefficient with the depth, $-4\alpha(f)(z-z_0)$.

Here, we will employ subscripts $r$ and $s$ to refer to the reference phantom and sample (tissue), respectively. In SDM, it is assumed macroscopic homogeneity within the ROI. Thus, backscattering can be simplified to be a function of frequency, i.e. $B_s(f,z) = B_s(f)$ and $B_r(f,z) = B_r(f)$. The transmit pulse for sample and phantom imaging is assumed to be identical, $P_s(f) = P_r(f)$. Moreover, assuming the speed of sound (SOS) in the reference phantom and sample to be equal while using the same imaging system for them, the diffraction terms are assumed to be equivalent, as $D_s(f,z) \sim D_r(f,z)$. Towards obtaining the attenuation estimation of the sample, $\boldsymbol{\alpha_s(f)}$, the ratio of power spectra of the sample to the reference phantom $RS(f,z)$, is defined as in equation (3)**Error! Reference source not found.** after applying the above assumptions and taking the natural logarithm of the two sides.

$$Ln\,[RS(f,z)] = Ln\left[\frac{B_s(f)A_{0,s}(f,z_0)}{B_r(f)A_{0,r}(f,z_0)}\right] - 4[\alpha_s(f) - \alpha_r(f)](z - z_0) \quad (3)$$

Equation (3) is a function of depth and frequency. At a given spectral component within the pulse $f'$, the first term on the right-side is a constant, allowing for the estimation of attenuation coefficient, $\alpha_s(f')$, through a linear loss modeling of natural log of power spectra ratio along the unit path length (corresponding to an exponential decay of a power spectral density component with distance). Over a range of usable frequencies, attenuation coefficient as a function of frequency is obtained. A proper frequency domain modeling for the attenuation behavior results in estimating the attenuation coefficient slope of the sample. It is noted that the units of attenuation coefficients, $\alpha_s(f)$ and $\alpha_r(f)$, in equation (3) are $Np/cm$.
Attenuation coefficient increases with frequency in soft biological tissues. In medical ultrasound, the attenuation coefficient in general is indicated to have a power-law type of behavior as $\alpha_s(f) = A.f^n$ with the power of $0.5 < n < 1.5$ [55-57], due to multiscale features of soft tissues. In practice, attenuation is typically approximated as a linear model of frequency, particularly for the frequency ranged of up to 12 MHz or up to 20 MHz [57], when the bandwidth is limited. However, some studies on high frequency attenuation estimation have still suggested the reasonable justification of using linear modeling over the power-law due to the higher variations of power-law fitting models even for repetitions done on single subjects. For instance, Raju et. al [58] made such conclusions from robustness of linear modeling over power-law when estimating *in vivo* attenuation of healthy forearm dermis in the range of 14 to 50 MHz. Reliable estimations of a power law model parameters require large bandwidths, otherwise over a narrow bandwidth, multiple power-law functions may produce almost identical attenuation curves. This could lead to unstable or non-physical fitting results with a large range of variation. Also, other than a larger bandwidth, such assessment of attenuations needs to be performed at several independent frequencies for reliable estimations. Thus, linear attenuation modeling often reasonably fits well over a limited bandwidth and is frequently employed. As a recent 2025 study, Omural et al. investigated high frequency characterization of lymph-node tumors in mice by adopting the linear frequency modeling (over the range of 7-30 MHz) and reported a decrease in attenuation with tumor growth [59].
The linear frequency modeling of the attenuation coefficient is either treated as a one-parameter model with an ACS of $\beta$ (equation (4)), or as a two-parameter model with the slope of $\alpha_1$ and an additional intercept of $\alpha_0$ (equation (5)).

$$\alpha(f) = \beta \cdot f \quad (4)$$
$$\alpha(f) = \alpha_1 \cdot f + \alpha_0 \quad (5)$$

In this study, we employed both models and compared their performance through a study on homogeneous tissue-mimicking phantoms with known acoustic properties.

## 3 Materials and Methods

### 3.1 Tissue-mimicking Phantoms

To verify the accuracy and robustness of the high-frequency attenuation estimation technique, we first implemented the SDM in customized tissue-mimicking phantoms with known acoustic properties (i.e. attenuation) prior to performing *in vivo* tissue analysis. Here, a high-attenuation phantom was used as the reference medium to estimate the ultrasound attenuation of a second phantom, both scanned using the same imaging setup as the *in vivo* acquisitions.
To enhance generality and robustness in this study, we employed two independently fabricated and calibrated phantoms from two laboratories by experts (co-authors T.L. and C.B.). These phantoms were scanned using the same imaging setup as the in vivo acquisitions, and the proposed method was applied consistently across both datasets. As summarized in Table 1, these phantoms, referred to as Ph # 1 and Ph # 2, have an ultrasound attenuation coefficient of 1.12 dB/cm.MHz and 0.7 dB/cm.MHz, respectively, as calibrated by the manufacturers.

**Table 1**. Characteristics of tissue-mimicking phantoms used for the evaluation of the attenuation estimation technique.

| **Tissue-mimicking Phantoms** | Ultrasound Attenuation Coefficient [dB/cm.MHz] | Manufacturer |
|---|---|---|
| Phantom 1 (Ph # 1) | 1.12 | Sun Nuclear Virginia, VA, USA. (Formerly CIRS, Inc.) |
| Phantom 2 (Ph # 2) | 0.7 | University of Wisconsin-Madison, Madison, WI, USA. |

### 3.2 Preclinical Study: Swine Periodontal Tissues

The pre-clinical swine model for assessing attenuation in periodontal tissues consisted of 10 Sinclair mini pigs. Pigs' oral tissues were scanned across all four oral quadrants (maxillary left, maxillary right, mandibular left, and mandibular right). Within each quadrant, oral soft tissues adjacent to four teeth at their interproximal sites were scanned. Depending on the reference tooth, the interproximal space is either mesial (closer to the face mid-line) or distal (farther from the face mid-line). The convention for defining a distal or mesial oral site between consecutive teeth are illustrated in Figure 2 (a). It is noted that when placing the toothbrush-sized ultrasound transducer, the distal oral site of a tooth corresponds to the mesial oral site of the adjacent tooth as the probe moves from the lip toward the oral cavity. For example, the scans of 'molar 1 - distal' and 'molar 2 – mesial' sites appear the same.  In Figure 2 (b), the transducer positioning for intraoral site imaging from the buccal (cheek) side is represented for a swine subject.
The five interproximal sites enrolled in this study are listed below which are also marked on the illustration in Figure 2 (a):

- Premolar 3 – Mesial (PM3-Mes),
- Premolar 3 – Distal (PM3-Dis),
- Premolar 4 – Distal (PM4-Dis),
- Molar 1– Distal, (M1-Dis), and
- Molar 2 – Distal (M2-Dis).

This preclinical study protocol was approved by the Institutional Animal Care and Use Committee at the University of Michigan (PRO00010333). The University of Michigan's Animal Care & Use Program is accredited by AAALAC International.

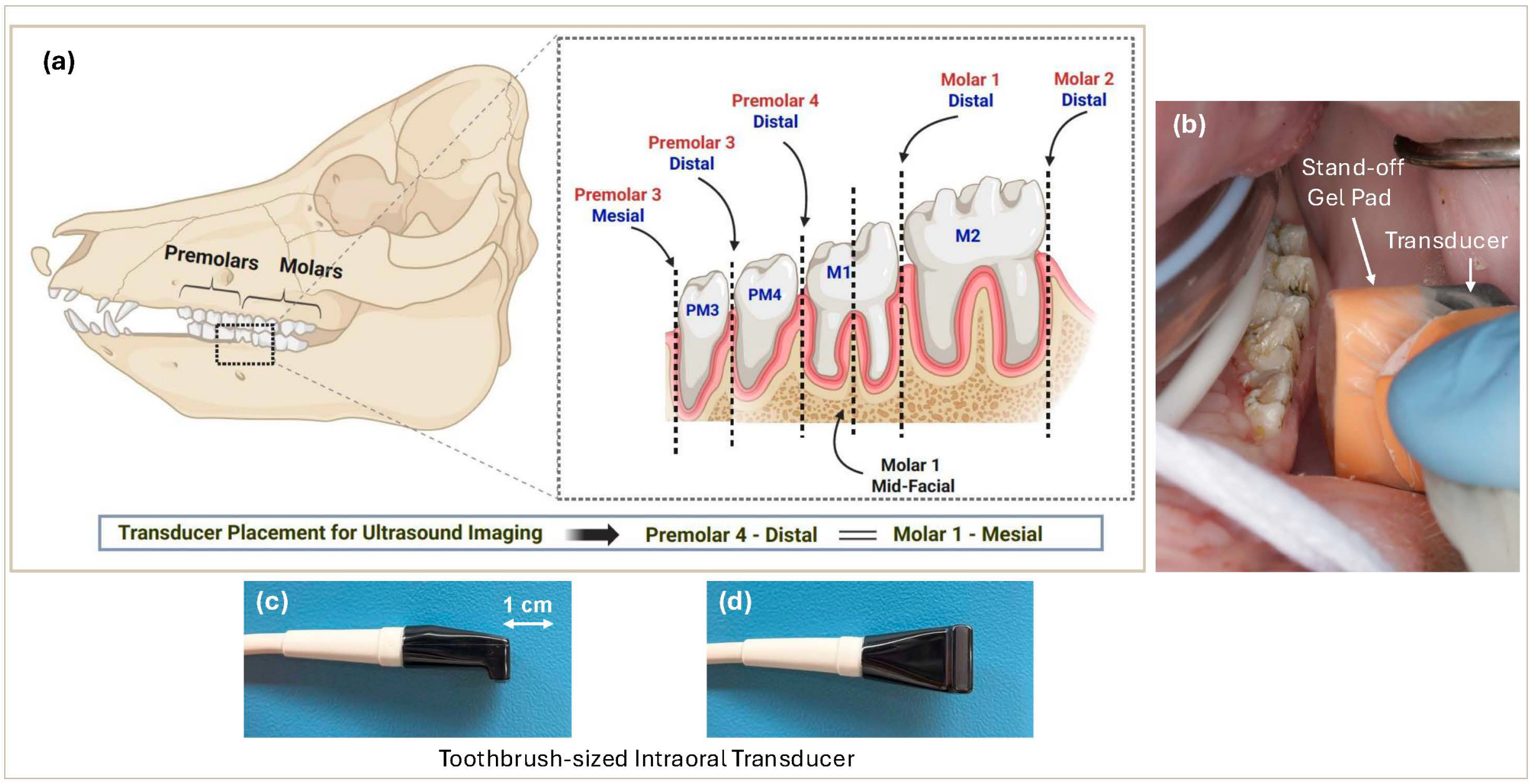

**Figure 2**. (a) Demonstration of interproximal and mid-facial oral sites as potential locations for ultrasound image acquisition in a swine case within the sagittal (i.e. left-right) plane. (b) Transducer positioning for intraoral site imaging from the buccal (cheek) side. Five enrolled oral sites in this study are also marked. (c) and (d): High-frequency toothbrush-sized transducer (linear-array) for intraoral ultrasound imaging shown from side and front views.

## 3.3 Ultrasound RF/IQ Data Acquisition: Phantoms and Tissues

For acquiring ultrasound IQ data (IQ: In-phase and Quadrature) in tissue-mimicking phantoms (TMP) and the pre-clinical swine model, a US FDA-approved clinical ultrasound imaging system (ZS3, Mindray Innovation Center of North America, San Jose, CA, USA) was employed paired with a toothbrush-sized linear array transducer with 128 elements (L30-8, Mindray Innovation Center of North America, San Jose, CA, USA). The transducer's side and front views are shown in Figure 2 (c) and (d), respectively. The transducer has a center frequency of 18 MHz and operated at the transmit/receive frequency of 24 MHz for both studies. The elevational focus of the transducer is fixed at the depth of 8 mm. The side-facing aperture dimension is 16.2 mm. IQ data were exported for post-processing purposes to estimate attenuation. RF data was obtained from IQ remodulation.

A stand-off gel pad (Aquaflex Ultrasound Gel Pad, Parker Laboratories, NJ, USA) was used to place desired oral tissues within the focal region of the transducer, as shown in Figure 2 (b). The gel was cut to match the dimension (height and width) of transducer surface with a thickness of about 5 mm. Further details on transducer preparation for intraoral imaging are provided in a recent publication from our group [60].

## 3.4 ROI Selection in Oral Tissues

For each swine scan, a manually selected rectangular ROI was placed within the gingival tissue to estimate the attenuation coefficient. The ROIs excluded local regions that incorporated artifacts such as clutter from neighboring teeth, any potential reverberation as well as epithelium and rete pegs within the oral scans. Moreover, some of the scans were excluded from the study due to one or more of the following factors: limited ROI size within the acquired imaging plane, macroscopically heterogeneous tissue composition within the imaging plane, presence of large imaging artifacts such as clutter and reverberation, or insufficient overall scan quality. These prevented placement of a suitable ROI for QUS analysis.

In 10 pigs, four oral quadrants in each pig were scanned, each quadrant with five distinct interproximal oral sites making a total of 200 scans. IQ data were unavailable for 24 of 200 oral sites, leaving 176 scans for attenuation assessment. Of these, 14 scans (8%) did not meet the requirements for attenuation ROI placement mentioned above and were thus excluded from our analysis. A total of 162 scans were eventually included for attenuation estimations.

## 3.5 Spectral Domain Analysis

In implementing the spectral difference method, the Fast Fourier transform was applied to segments of RF data gated by a Hanning window. The gate size choice reflects a trade-off between the sufficient stability of the power spectral representation of the gated signal and the stationary requirement in each gated region (being locally homogeneous) for the attenuation estimation. The gate size of approximately 8 times of the measured transmit pulse was adopted for both tissue-mimicking phantoms and for *in vivo* periodontal tissue analyses. It was selected consistent with a suggested value in literature [54] and also via the evaluation of the center frequency and the-full-width-half-maximum (FWHM) of the power spectrum of the gated signal in a simulation study. A gate overlap ratio of 50% was used to maintain axial resolution [54, 61]. Power spectra went through frequency smoothening to reduce spectral noise artifacts.

While biological tissues are three-dimensional by nature, so is the variation of acoustical parameters characterizing them such as density, compressibility, and absorption. Many QUS estimation approaches such as SDM for attenuation rely on a simplified analysis of a 3D scattering field. As the tissue frequency response fluctuates spatially, frequency dependent attenuation estimations require statistical averaging [62]. Thus, lateral averaging was performed across consecutive scanlines (with an increment of three scanlines) at each gate depth within each ROI to mitigate the effect of unavoidable speckle noises.

## 3.6 Fine-tuning Diffraction Correction in Tissue Imaging

Diffraction, or beam spreading, is an inherently depth-dependent phenomenon. During *in vivo* tissue imaging, the starting depth of tissue within the imaging plane can vary due to operator-dependency as well as the practical constraints imposed by the amount of intraoral space available. To ensure a more consistent diffraction correction of tissue scans, multiple phantom scans were acquired with an incremental change of 0.5 mm in the imaging start depth, as shown in Figure 3. For each tissue scan, the phantom scan with the closest matching start depth was selected. Since speed of sound (SOS) plays a crucial role in the diffraction phenomenon, this approach fine-tunes any diffraction-related discrepancies arising from a difference in SOS between the coupling gel and the soft tissue. Therefore, diffraction- and SOS-related variations across acquisitions may be minimized.

## 3.7 Statistical Method

For statistical significance analysis, the normality of data was assessed using the Shapiro-Wilk test along with the Q-Q plot visualization and $R^2$ of a linear correlation. A one-way ANOVA followed by a Tukey-Kramer post-hoc test was employed for significance test and adjusted p-values were reported for pairwise comparisons.

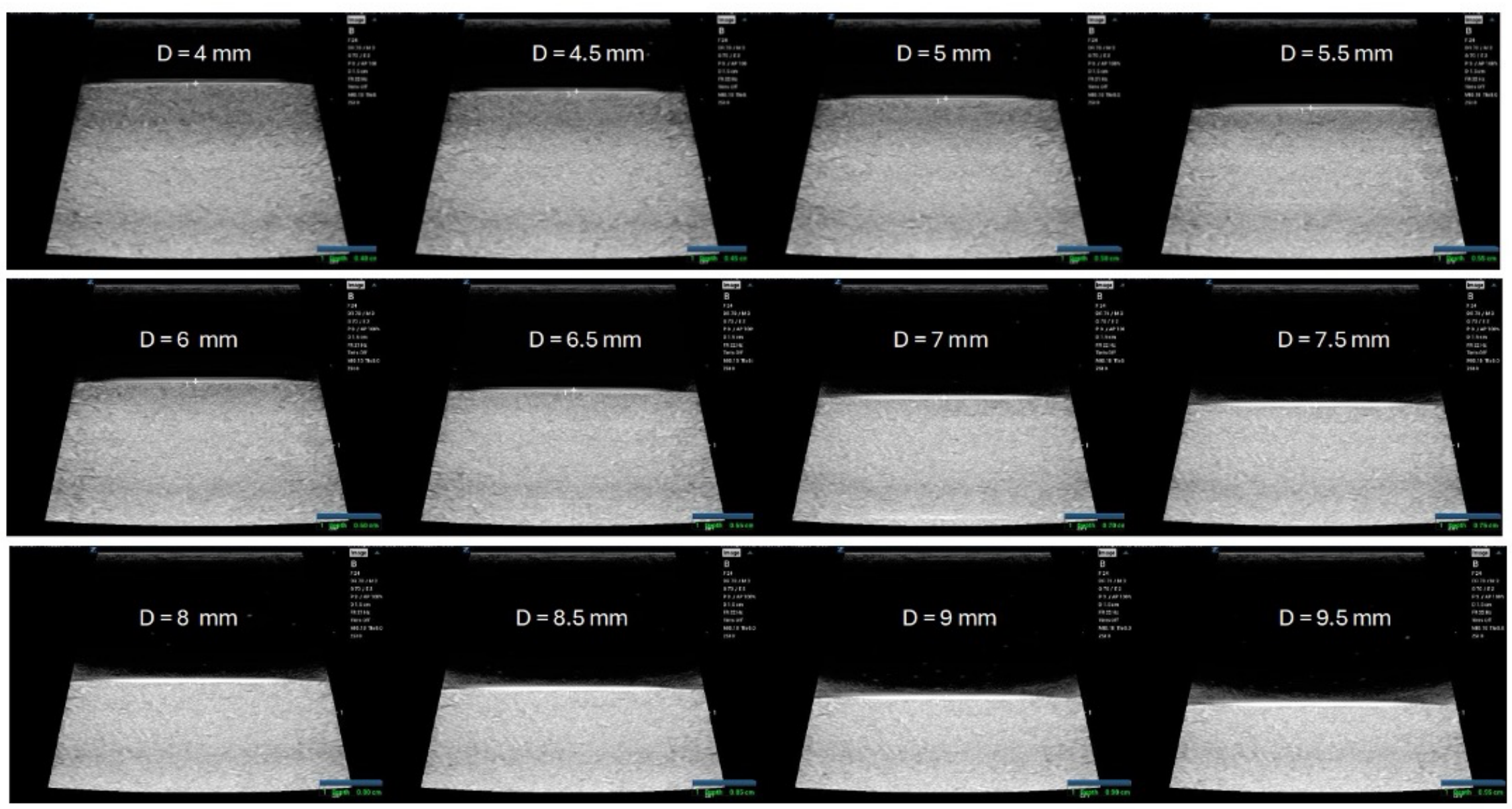

**Figure 3**. Ultrasound images of phantom 2 (Ph # 2) acquired at different imaging start depths, ranging from 4 mm to 9.5 mm with an increment of 0.5 mm using the same preset.

# 4 Results

## 4.1 Attenuation of Tissue-mimicking Phantoms

For the phantom study, 15 rectangular ROIs with random sizes were analyzed within the target phantom (Ph # 2). Figure 4 shows two representative ROI selections (left column) and their corresponding attenuation estimations using the one-parameter (middle column) and two-parameter (right column) frequency modeling. The estimated attenuation coefficient slopes, $\beta$ and $\alpha_1$ and the intercept $\alpha_0$ are reported on each attenuation-vs-frequency plot.

Figure 5 summarizes the attenuation coefficients from all 15 ROIs as boxplots to assess estimation robustness for both frequency modeling approaches. The mean attenuation coefficient using the one-parameter model ($\beta$) was 0.84 $\pm$ 0.17 dB/cm.MHz and for the two-parameter model ($\alpha_1$) was 0.83 $\pm$ 0.37 dB/cm.MHz. Considering the manufacture-calibrated attenuation coefficient for Ph # 2 was 0.7 dB/cm.MHz, both approaches provided estimations within 20% of this value, particularly at such a high frequency. The estimations from these two approaches were compared based on a paired student's t-test with their normality confirmed using the Shapiro-Wilk test applied on their differences. The two sets of estimations were not statistically significant from each other (p-value >0.05). While their means were very close, the lower standard deviation of $\beta$ compared to $\alpha_1$ indicated a more robust estimator when the one-parameter model was used.

Additionally, the preference for the one-parameter model is also physically more justified since its attenuation approaches zero as frequency goes to zero. The two-parameter model, despite its added flexibility by introducing a second parameter, resulted in a non-zero attenuation at zero frequency, which is unphysical. Therefore, one-parameter modeling was chosen over the two-parameter approach.

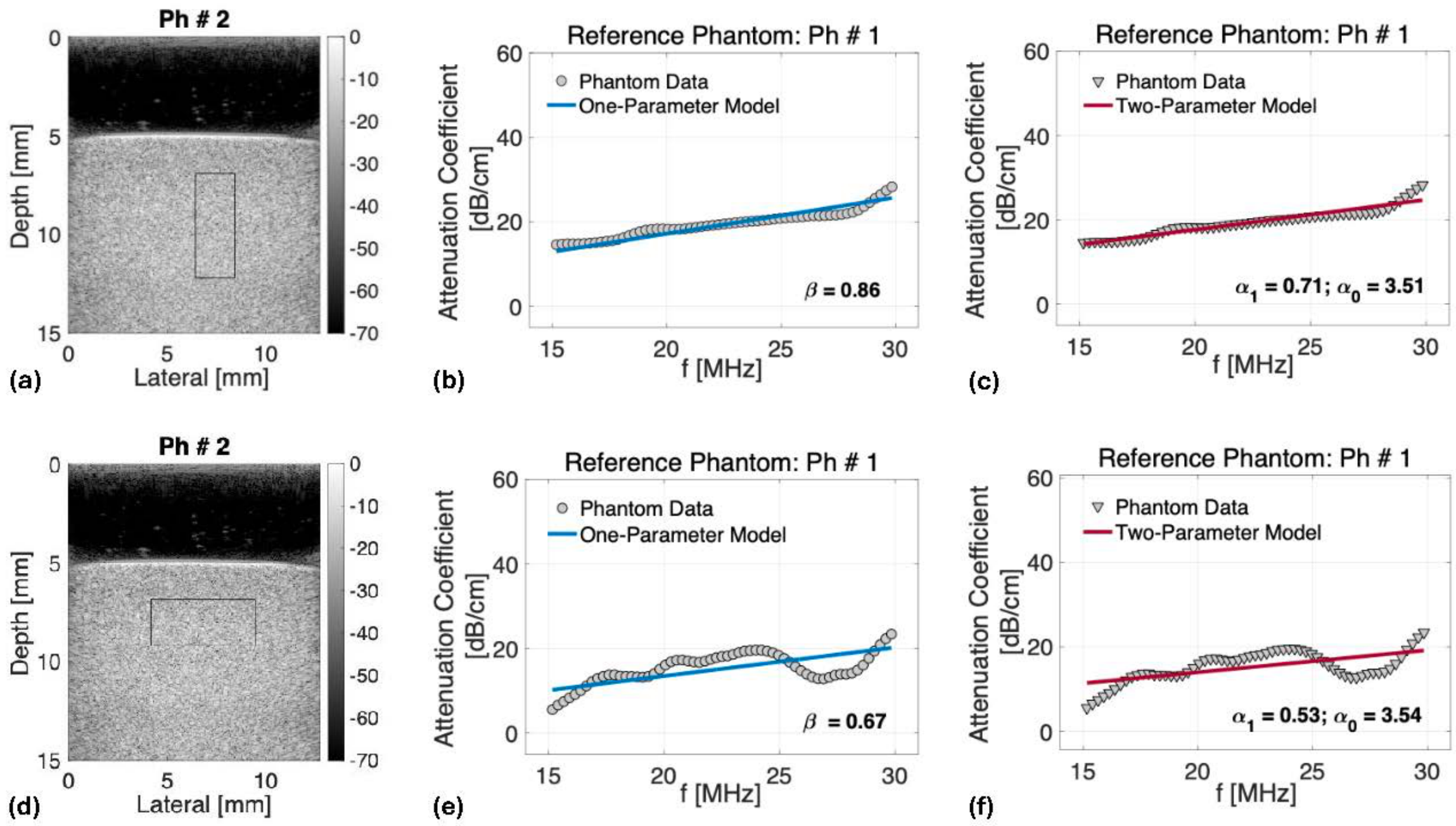

**Figure 4**. Two representative regions of interest (ROI) selections for attenuation estimation in the tissue-mimicking phantom 2 (Ph # 2) using Ph # 1 as the reference medium. Left column (a and b): B-mode images with example ROI (rectangular box); middle column (b and e): one-parameter attenuation estimation, right column (c and f): two-parameter attenuation estimation. The estimated attenuation coefficient slopes are reported for each plot.

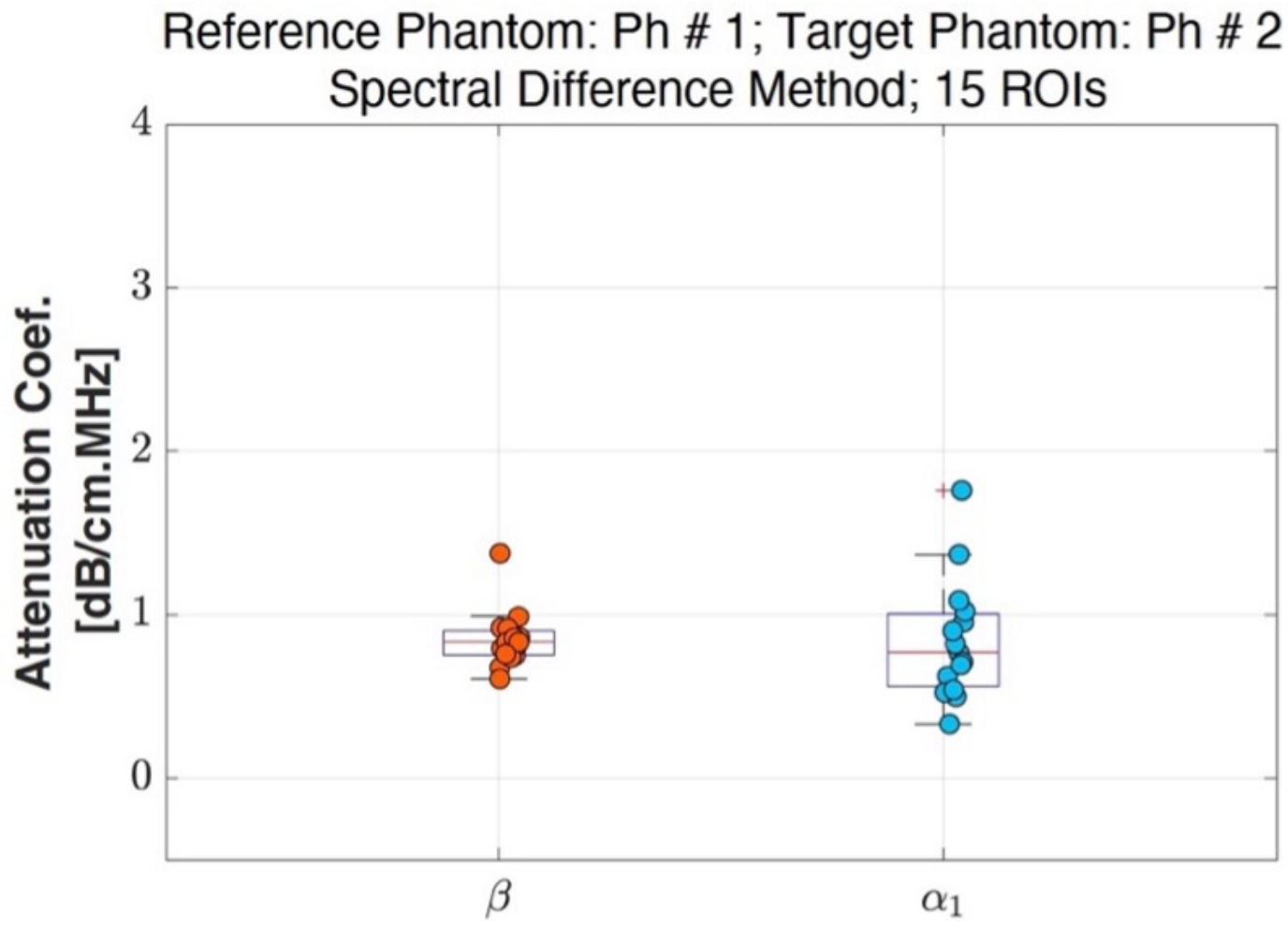

**Figure 5**. Summary of attenuation estimations of tissue-mimicking phantom 2 (Ph # 2) across 15 different random ROIs using Ph # 1 as the reference media to assess the estimation robustness. Boxplots compare

attenuation estimation results from one-parameter ($\beta$) and two-parameter ($\alpha_1$) frequency modeling approaches for identical ROIs.

## 4.2 Periodontal Tissues

### 4.2.1 Ultrasound Images of Swine Oral Tissue

Figure 6 presents four high-resolution ultrasound B-mode images of oral tissues at four oral quadrants of a swine subject, showcasing oral landmark structures. These images are acquired intraorally at the interproximal spaces between the first and second molars from the buccal (cheek) side. Panels (a) and (b) correspond to the maxillary (upper jaw) right and left quadrants, respectively. Panels (c) and (d) display the corresponding mandibular (lower jaw) right and left quadrants. It is noteworthy that in dentistry, the convention for assigning the left and right directions in identifying oral quadrants are based on the subject's left and right sides, as illustrated by the four quadrants at the center of Figure 6.

Some of the important oral landmarks on a B-mode image are annotated in Figure 6 (b), namely bone, epithelium layer, attached gingiva and lining mucosa. The keratinized epithelium and the bone surface both appear as thin and curved hyperechoic layers on ultrasound B-mode images. Gingival and mucosal tissues are shown as granular typical speckle pattern of ultrasound B-mode images. It is noted that these images are recorded using the compounding second harmonic imaging (CSH24) mode of the scanner at transmit/receive frequency of 12/24 MHz which provides a superb visual representation of swine oral tissues with better image quality and reduced noise compared to used images derived from the in-phase and quadrature (IQ) data processing. However, IQ data analysis is mandatory since the complex processing chain for clinical quality images does not allow for attenuation analysis.

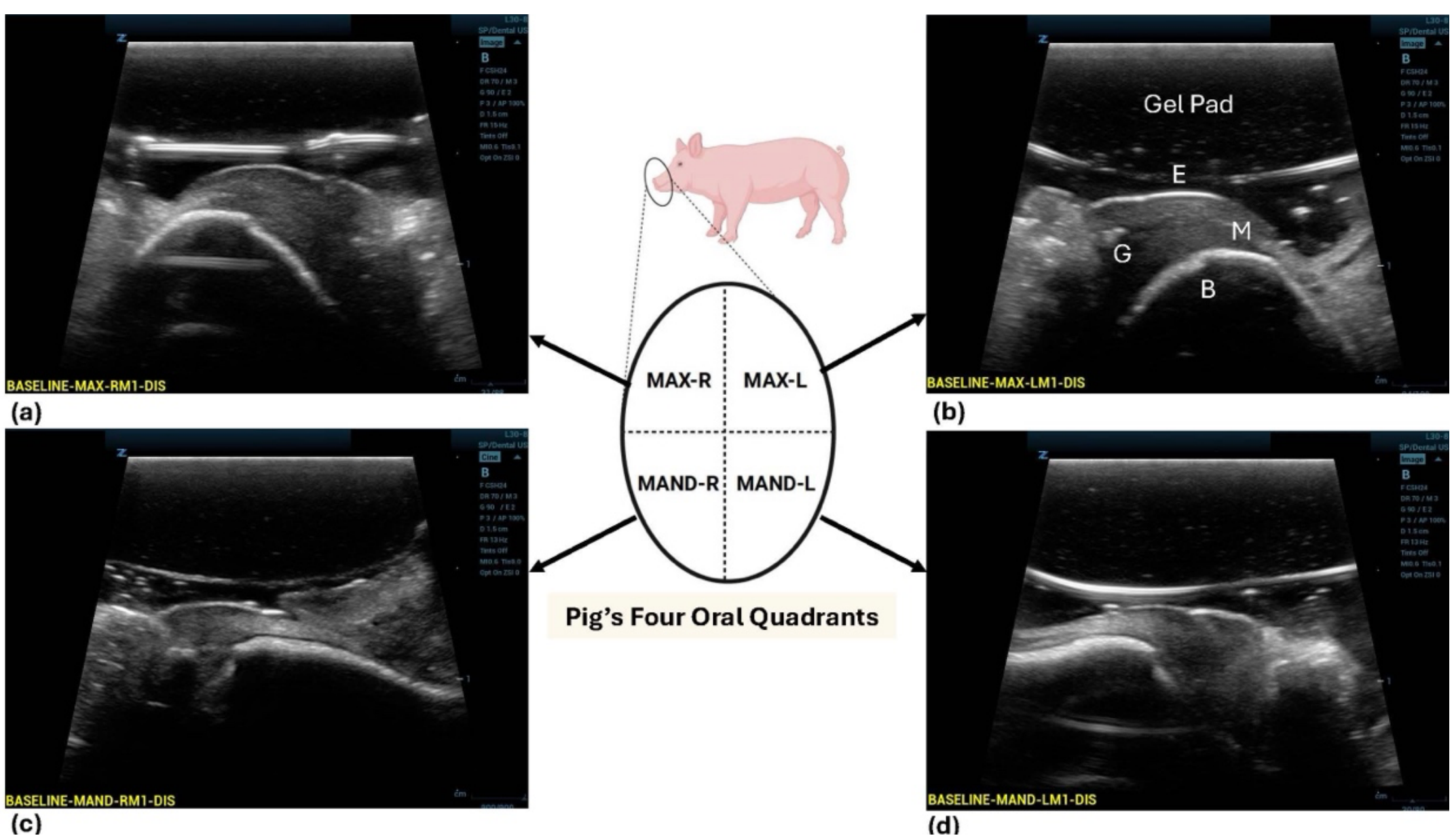

**Figure 6**. Ultrasound B-mode images of periodontal tissues at interproximal oral sites between the first molar and second molar (M1-Distal) from all four oral quadrants (a) right maxilla, (b) left maxilla, (c) right mandible, and (d) left mandible. Important oral landmarks are annotated in (b): B: bone; G: interdental

gingiva; M: lining mucosa; E: epithelium Layer). Note: B-mode images presented here are recorded by employing CSH24 mode of the ultrasound scanner.

### 4.2.2 Attenuation of Swine Oral Tissue

Attenuation coefficient estimations of interdental gingival tissues in two swine subjects are presented in Figure 7. These are scans from interproximal (distal) areas of the first molar. The upper row represents this oral site at the left maxillary quadrant and lower row shows that of right maxillary quadrant. The estimations are based on the one-parameter frequency model. Here, we compare estimations when each of the two phantoms were set as the reference medium. The left panel displays B-mode images reconstructed from IQ data with ROIs annotated as white rectangles. Attenuation coefficient estimates of phantom 1 (Ph # 1) and phantom 2 (Ph #2) are shown in panels (b)/(e), and panels (c)/(f), respectively. The derived attenuation coefficients for these subjects are:

- Subject # 7: $\beta_{(Ph\ \#1)}$= 1.13 dB/cm·MHz; $\beta_{(Ph\ \#2)}$= 1.19 dB/cm·MHz
- Subject # 9: $\beta_{(Ph\ \#1)}$= 0.75 dB/cm·MHz; $\beta_{(Ph\ \#2)}$= 0.78 dB/cm·MHz

Considering the independent fabrication of the two reference phantoms, variations in attenuation estimations are expected. As an example, Figure 8 (a) presents boxplots comparing the in vivo attenuation coefficient estimations at the distal side of the second molar (M2-Dis) using both phantoms. The average of attenuation coefficient slope from Ph # 1 was 1.22 $\pm$ 0.34 dB/MHz.cm and for Ph # 2 was 1.13 $\pm$ 0.32 dB/MHz.cm. Their variability is compared using Bland-Altman analysis as shown in Figure 8 (b). Here, the solid black line represents the bias, and the two dotted-dashed red lines show lines of agreements, located at a distance of $1.96.\sigma(Difference)$ around the bias, with $\sigma$ being the standard deviation of the difference between $\beta_{(Ph\ \#1)}$ and $\beta_{(Ph\ \#2)}$. It is observed that the bias of the attenuation coefficient difference among estimation pairs are relatively low, (Bias = 0.099 dB/MHz.cm) with no estimates located beyond the two agreement lines around the bias.

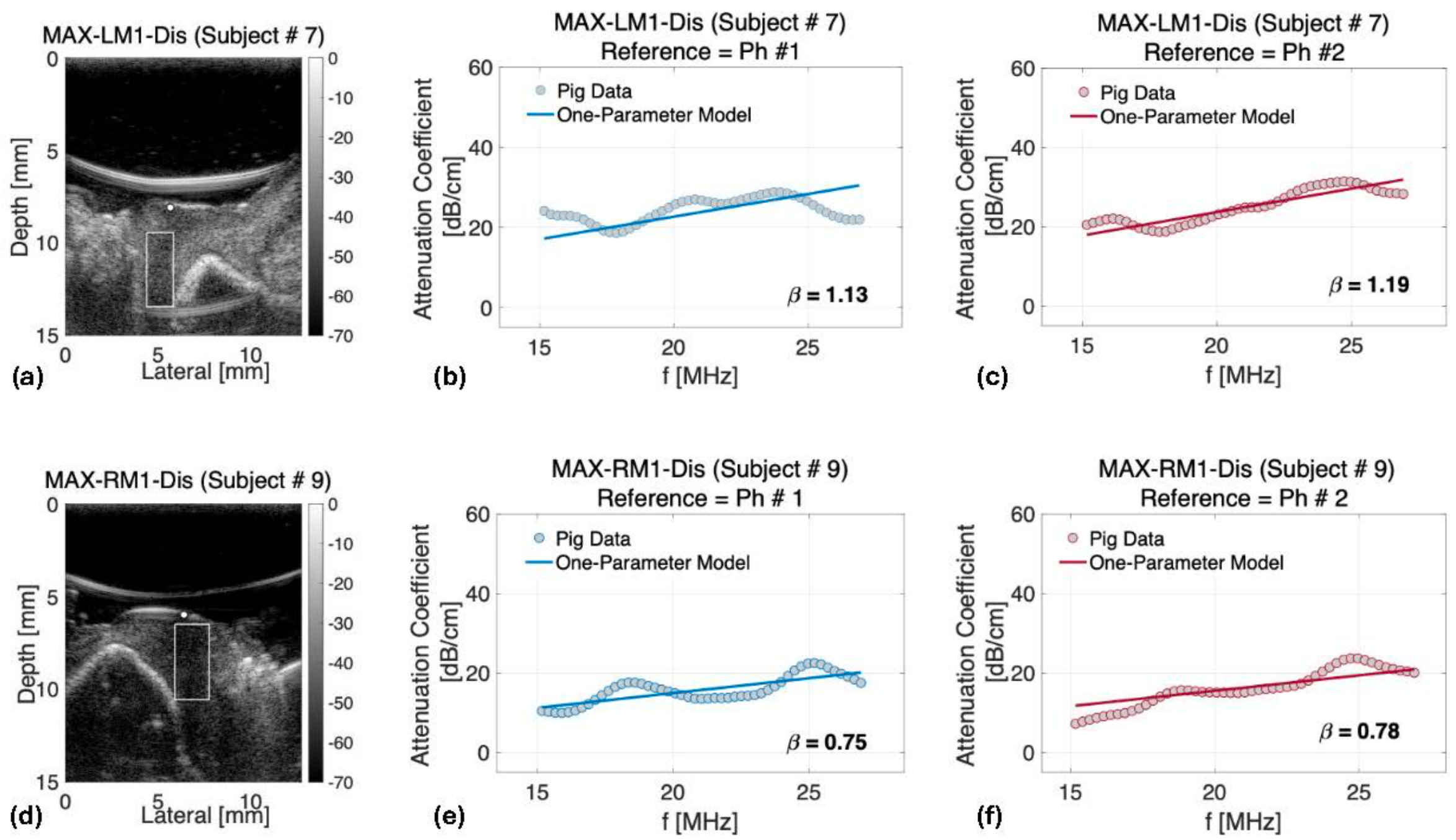

**Figure 7**. Two examples of attenuation coefficient estimation in gingival tissues in two subjects at the interproximal (distal) oral site around molar 1 using the one-parameter model. The effect of the phantoms Ph # 1 and Ph # 2 as the reference media is compared. Left Column (a and d) : B-mode images with the regions of interest shown as white boxes for attenuation estimation. Middle Column (b and e): attenuation coefficient as a function of frequency when Ph # 1 was the reference medium. Right Column (c and f): attenuation coefficient using Ph # 2 as the reference medium. Top row exhibits a case from the right maxillary oral quadrant and bottom rows shows a case in the left maxillary quadrant.

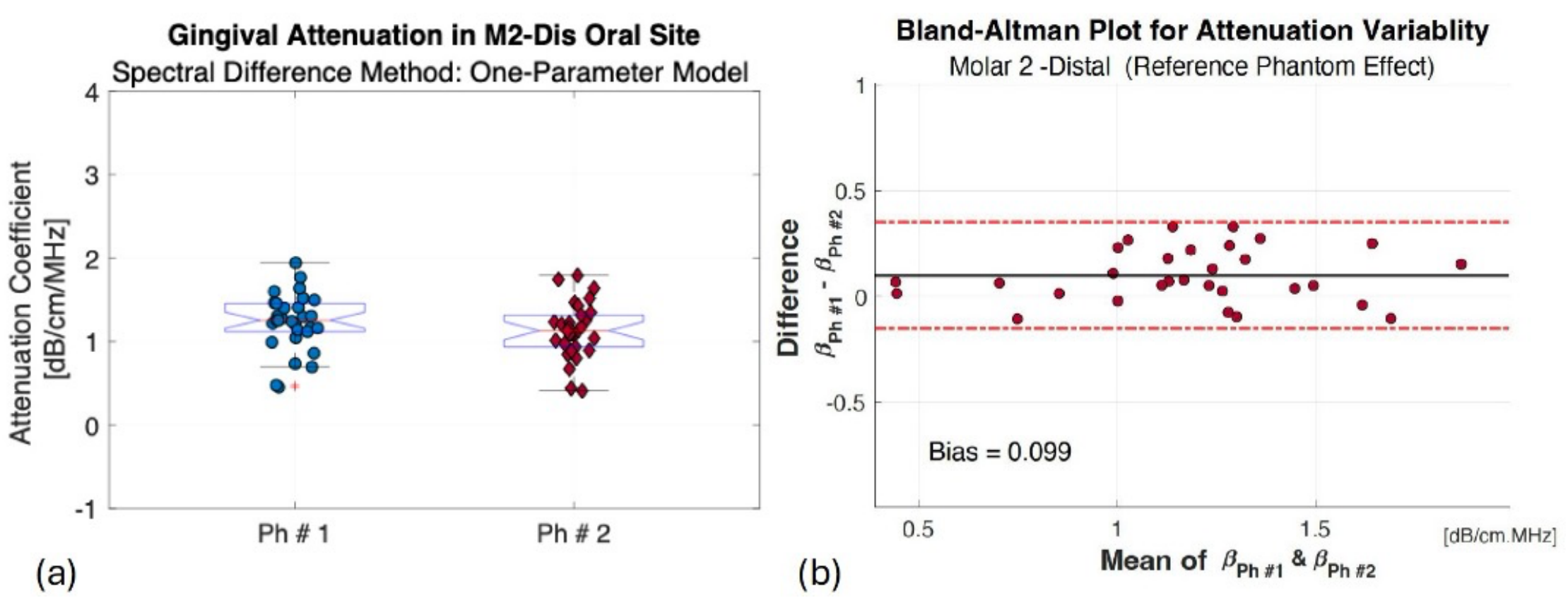

**Figure 8**. (a) Summary of attenuation coefficient estimation ($\beta$) of the interdental gingival tissues at the distal site of second molars using both reference media (Ph # 1 and Ph # 2) using the one-parameter attenuation model. (b) The Bland-Altman plot comparing the variability of estimates ($\beta$) using two phantoms. The black solid line shows the bias (=0.099 dB/cm·MHz) and the dotted-dashed red lines show limit-of-agreement lines around the bias located at $bias\ \pm 1.96.\sigma(Difference)$.

For oral tissue analysis, we chose Ph # 2 as the reference medium due to its smaller bead size and its larger physical size to minimalize wall reflection effects.

For a holistic characterization of ultrasound ACS in healthy periodontal tissues, a total of five oral sites were investigated, and a summary is presented as a boxplot for each oral site in Figure 9 with the individual measurements as scattered plots. From left to right, the boxplot represents PM3-Mesial, PM3-Distal, PM4-Distal, M1-Dis and, M2-Distal, arranged from anterior to posterior within the oral cavity.

Using the one-parameter frequency modeling, the mean and the median (Q1 | Q3) ultrasound ACS for these fives oral sites are reported in Table 2.

Interdental gingival tissues at the mesial and distal third premolar (PM3-Mes and PM3-Dis) showed an elevated mean attenuation compared to their posterior sites with the PM3-Mes having the highest mean attenuation coefficient estimates. The two molar interdental gingival tissues (M1-Dis and M2-Dis) exhibited similar mean attenuation coefficients of 1.18 and 1.13 dB/MHz.cm, respectively.

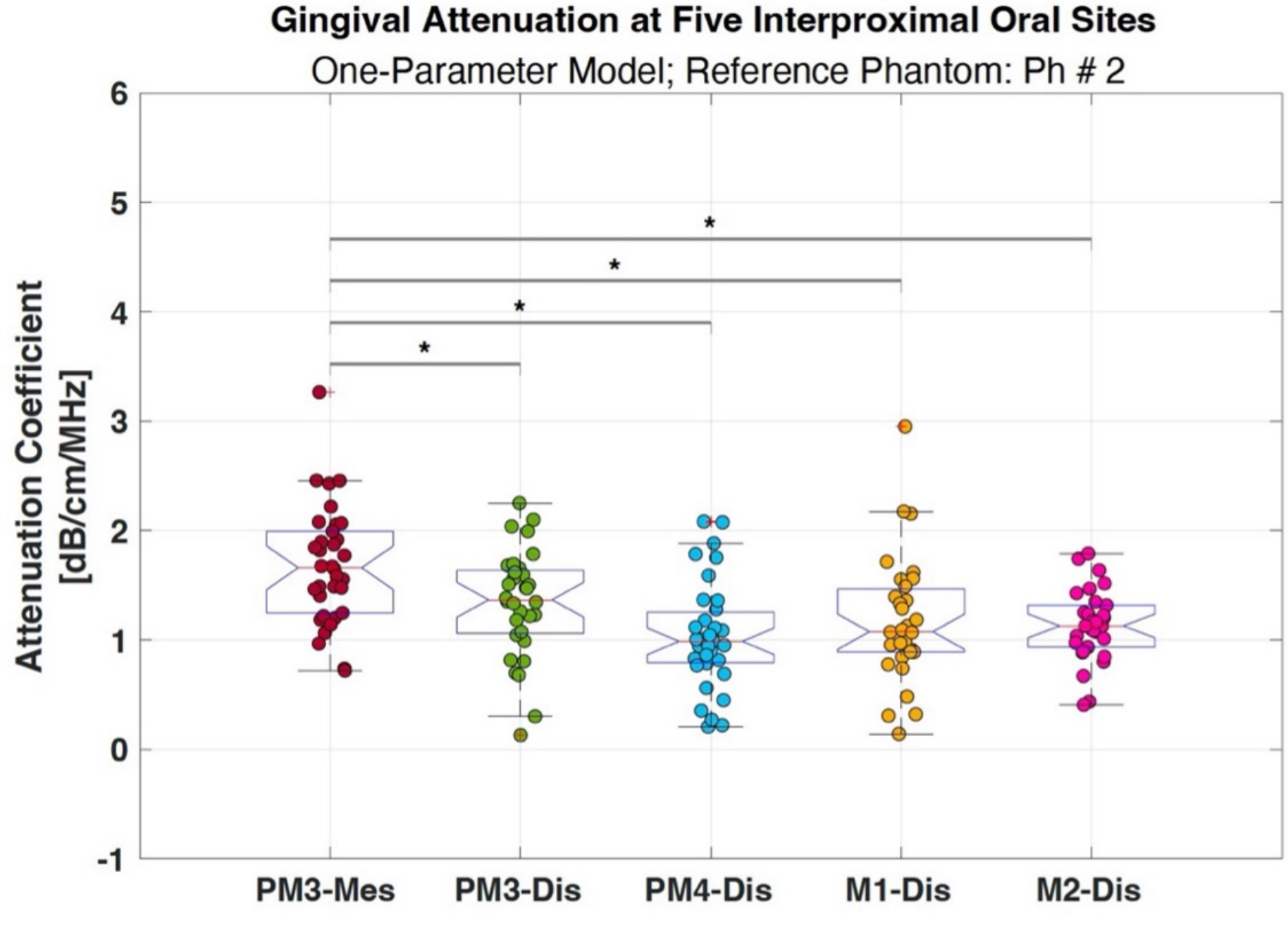

**Figure 9**. Quantification of ultrasound attenuation coefficient of the interdental gingival tissues across the swine population incorporating all four oral quadrants using the one-parameter frequency modeling ($\beta$). These oral sites are premolar 3-mesial, premolar 3 distal, premolar 4 distal, first molar distal and second molar distal. Asterisks indicate statistically significant pairs ($p < 0.05$).

**Table 2**. Ultrasound attenuation coefficients of the interdental gingiva in a swine cohort at fives oral sites.

| Interproximal space | Attenuation Coefficient [dB/MHz.cm] | |
|---|---|---|
| | **Mean** | **Median (Q1\|Q3)** |
| PM3-Mesial | 1.68 | 1.66 (1.25 \| 1.99) |
| PM3-Dis | 1.34 | 1.37 (1.6 \| 1.64) |
| PM4-Dis | 1.03 | 0.99 (0.80 \|1.25) |
| M1-Dis | 1.18 | 1.08 (0.89 \| 1.47) |
| M2-Dis | 1.13 | 1.28 (0.93 \| 1.32) |

To assess whether the characteristic acoustic attenuation coefficients differed, i.e. if they are statistically distinguishable across oral sites, a one-way ANOVA followed by a Tukey-Kramer post-hoc test was performed. Pairwise comparisons with their adjusted p-values are reported in Table 3. Normality of the attenuation coefficients within each oral site was evaluated using the Shapiro-Wilk test, which indicated that the data were largely consistent with a normal distribution. This assessment was further supported by Q-Q plot visualizations and analysis of correlation $R^2$, which demonstrated reasonable linearity by a $R^2$ of higher than 90%. Given these findings and the moderate sample sizes, the use of a parametric ANOVA-based approach was considered appropriate.

Multiple comparison analysis revealed that the gingival tissues at PM3-Mes exhibited significantly higher ACS compared to the other four sites. Specifically, the adjusted p-values for the comparison between PM3-Mes and its adjacent site, i.e. PM3-Dis, was $p < 0.05$ while between PM3-Mes and the remaining other oral sites involving molars yielded smaller p-values ($p < 0.001$). No statistically significant differences were observed among ACSs for any other pairwise comparisons.

Across all these five sites, the average ultrasound attenuation coefficient of gingival tissues was 1.27 dB/MHz.cm with a standard deviation of 0.54 dB/MHz.cm. When only the four distal sites were considered, the mean and standard deviation of ultrasound attenuation coefficient decreased to 1.17 dB/MHz.cm and 0.48 dB/MHz.cm, respectively.

**Table 3**. Pairwise statistical comparison of ultrasound attenuation coefficients of gingival tissues among five interproximal oral sites across the swine cohort. Asterisks indicate statistical significance ($p < 0.05$). (NS: Not Significant)

| **Pair-wise Comparison of Oral Sites** | | **Mean Difference** | **Adjusted p-values** | **Significance** |
|---|---|---|---|---|
| **PM3-Mes** | **PM3-Dis** | 0.34 | 0.03852 | (*) |
| **PM3-Mes** | **PM4-Dis** | 0.64 | 5.19E-07 | (*) |
| **PM3-Mes** | **M1-Dis** | 0.50 | 3.64E-04 | (*) |
| **PM3-Mes** | **M2-Dis** | 0.55 | 8.47E-05 | (*) |
| PM3-Dis | PM4-Dis | 0.30 | 0.0865 | NS |
| PM3-Dis | M1-Dis | 0.16 | 0.6889 | NS |
| PM3-Dis | M2-Dis | 0.21 | 0.4649 | NS |
| PM4-Dis | M1-Dis | -0.14 | 0.7711 | NS |
| PM4-Dis | M2-Dis | -0.09 | 0.9336 | NS |
| M1-Dis | M2-Dis | 0.04 | 0.9966 | NS |

# 5 Discussion

This study, to our knowledge, serves as the first quantification of ultrasound attenuation of oral soft tissues in a preclinical study using standard techniques in physical acoustics and supporting validations from industry-standard tissue mimicking phantoms. This is a significant step towards advancing QUS imaging technologies in dental healthcare. Our investigation involved a comprehensive analysis of several oral sites including two molars and two premolars using the spectral difference method.

Gingival tissues are the first line in inflammation onset due to food residue from their proximity to the pocket (sulcus) with potential bacterial accumulation (Figure 1) Advanced periodontal diseases such as periodontitis and the associated potential tooth loss can simply start from an early gingival inflammation. Thus, an early inflammation diagnosis that may trigger any preventive intervention is possible through an accurate assessment of gingival tissue health. Towards advancing QUS imaging with the potential to offer a quantitative and non-invasive oral inflammation biomarker, the ultrasound attenuation coefficient is a crucial tissue parameter that characterizes the acoustic properties of oral tissues. Moreover, its quantification is a significant confounding parameter for other QUS techniques which rely on the knowledge of ultrasound attenuation to compensate radiofrequency data before any further analysis.

The accuracy of the high-frequency attenuation estimation techniques implemented in this study was thoroughly validated through our measurements on tissue-mimicking phantoms with a (manufacturer) measured attenuation coefficient of 0.64 dB/cm.MHz at 10 MHz and predicted as 0.67 dB/cm.MHz at 25 MHz. The phantom favorably compared to the literature [63]. The implemented spectral difference method yielded a mean attenuation coefficient of 0.84$\pm$0.17 dB/cm.MHz, which includes the predicted value within the 95%-confidence interval. It is notable that our ACS estimation for Ph # 2 (fabricated and calibrated at the University of Wisconsin) was based on employing a reference phantom (Ph # 1) fabricated and calibrated at CIRC Inc. High-frequency attenuation may differ between these phantoms and could lead to error. Either or both phantoms could deviate from their here assumed slightly super-linear increase with frequency, i.e., proportional to $f^x$, with $x \approx 1.1$. Future work should include single-element transducer-based calibration of both employed phantoms at the imaging frequency of 24 MHz. However, such studies require dedicated hardware that was not available for the present study.

In an analysis of multiple interproximal oral sites of premolars and molars across a swine cohort of 10 (total of 162 scans), we showed a reproducabe attenuation estimate for gingival oral tissues with an average ultrasound attenuation of the four distal oral sites being 1.17 dB/MHz.cm with a standard deviation of 0.49 dB/MHz.cm.

The comparison across the five oral sites showed that premolar 3 – mesial (PM3 – Mes) site exhibited statistically significant and higher attenuation coefficient values than the other four oral sites. Notably, PM3 – Mes had overall smaller gingival tissue regions available for ROI placement in attenuation estimation among evaluated oral sites. Considering the presence of various fibrous structures within gingival tissue, known to provide support for teeth and roots against cyclic and non-cyclic external mechanical forces from mastication (chewing), see the first figure in [8], presence of such fibrous structures within the relatively confined region of PM3 – Mes may contribute to the elevated attenuation coefficient estimated for this site.

To provide context for these gingival tissue attenuation values, we compare our results with the ultrasound attenuation coefficients for other biological tissues in the literature. Normal liver tissue has been reported to have attenuation coefficients in the range of 0.4-0.55 dB/cm.MHz [64-67]. For breast fat (suggested as a reference for comparative studies with breast tumors), an attenuation coefficient of 0.73 $\pm$ 0.23 dB/cm.MHz has been reported [40]. In skeletal muscle, the attenuation coefficient, measured in fresh bovine tissues, ranged from 1.1$\pm$0.15 perpendicular to the muscle fibers and 2.9$\pm$0.23 dB/cm.MHz parallel to them, reflecting a directional dependence [68]. As an example of hard tissues, trabecular bone samples from human specimens exhibited a substantially higher attenuation coefficient, with a reported mean of 10.7 dB/cm.MHz and a standard deviation of 5.7 dB/cm.MHz [69].

Overall, the ultrasound attenuation for gingival tissues falls within the range reported for soft and structurally heterogeneous biological tissues, such as skeletal muscles, and higher than those tissues with relatively homogeneous structures such as livers. These measurements suggest that gingival tissues cause substantial ultrasound energy loss, likely reflecting its complex microstructure such as presence of various fiber alignments [8]. One potential contributor behind the larger standard deviation across the cohort can be the variability in ROI size used for attenuation estimation, largely dictated by the physical size of the available tissue within each scan. In addition, the aforementioned fibers, analogous to muscle, could cause a larger variability.
Between the two frequency-response modeling approaches evaluated for attenuation estimation, the one-parameter model outperformed the two-parameter one by providing estimates with smaller standard deviations in the phantom study. It is noteworthy that since the two tissue-mimicking phantoms were independently fabricated in different laboratories, the reasonable attenuation accuracy further demonstrate the generality and robustness of the implemented technique at this high frequency attenuation study.
Attenuation-frequency relationships in tissue-mimicking phantoms and biological tissues, especially at such high frequencies, may deviate from strict linearity due to inherent microstructural heterogenicity as well as unavoidable imaging system bias. Particularly, in implementing the spectral difference method using a reference phantom, a fundamental assumption is the uniform distribution of scatterers within the phantom relative to the resolution of the imaging system which often are violated to some degrees. Thus, ultrasound ACS should be interpreted as capturing the overall increasing trend of attenuation with frequency rather than a perfectly linear relationship. At the high imaging frequency used in this study, the increased variance observed in attenuation estimates of the phantom may partially reflect deviations from scatterer uniformity. Future work could systematically evaluate the impact of bead non-uniformity on attenuation estimation accuracy. Given the limited space for intraoral scanning and the limited size of oral tissues, extending our analysis of attenuation to the lining mucosa or assessing tissue anisotropy behavior on attenuation through multiple-angle beam steering remains challenging. A possible limitation of this study is the number of subjects and future studies with larger cohorts are encouraged. Additionally, further investigations are suggested to see whether there is any clinical significance of the elevated ACS of PM3-Mes relative to other oral sites. Future studies should also explore the ACS of the lingual side and providing a pairwise comparison of lingual vs. buccal sides at each oral site. Moreover, the effect of pathological conditions such as impact of periodontal inflammation on ACS variations is an important research direction as well as efforts on translating research to clinical studies. Evaluating QUS measurement variabilities through inter- and intra-operator analysis is also important in future studies to ensure estimation robustness.” Additionally, investigating non-linear power-law attenuation models represent another direction to be explored.

# 6 Conclusion

In conclusion, this study is to our knowledge the first to quantify the ultrasound attenuation in oral tissues, a significant acoustic property that both reflects intrinsic tissue characteristics and serves as a foundation for many QUS-based tissue characterization methods. We investigated high frequency gingival tissue attenuation through a comprehensive analysis of multiple interproximal oral sites premolar mesial to second molar distal across all four oral quadrants in a cohort of pigs imaged at 24 MHz. Gingival attenuation exhibited an attenuation coefficient that fell between reported attenuation values for breast fat and skeletal muscles and well above liver attenuation. Future studies will extend this framework to pathological conditions and in clinical studies on human oral tissue characterization such as oral inflammation assessment, healing monitoring in periodontal regenerative surgeries.

## Author Contributions

- D.P.: Conceptualization, Data Curation, Formal Analysis, Investigation, Methodology, Software, Validation, Visualization, Project Administration, Funding Acquisition, Writing Original Draft and Revision.
- A.S., A.R.B., C.Q., T.L., C.B., J.B.F.: Investigation, Review and Editing.
- H-L.C.: Conceptualization, Investigation, Methodology, Resources, Review and Editing, Funding Acquisition, Supervision.
- O.D.K.: Conceptualization, Data Curation, Investigation, Methodology, Project Administration, Resources, Review and Editing, Funding Acquisition, Supervision.

## Acknowledgment

This work was supported by the National Institutes of Health through award numbers R21DE029005 (PIs: H-L.C. and O.D.K.) and F32DE034986 (PI: D.P.).

## Conflict of Interest Statement

The authors declare that they do not have any conflict of interest associated with this work.

## Data Availability statement

The data supporting the findings of this study are available from the corresponding authors upon reasonable request.